\documentclass[12pt,preprint]{aastex}

\shorttitle{${\it XMM}$-${\it Newton}$ AND ${\it Chandra}$ OBSERVATIONS OF M~31}
\shortauthors{TAKAHASHI ET AL.}

\begin{document}


\title{${\it XMM}$-${\it Newton}$ AND ${\it Chandra}$ OBSERVATIONS\\ OF THE CENTRAL REGION OF M~31}


\author{HIROMITSU TAKAHASHI, YUU OKADA, MOTOHIDE KOKUBUN, AND KAZUO MAKISHIMA}
\affil{Department of Physics, The University of Tokyo,
7-3-1 Hongo, Bunkyo-ku, Tokyo 113-0033}
\email{hirotaka@amalthea.phys.s.u-tokyo.ac.jp}



\begin{abstract}

The archival ${\it XMM}$-${\it Newton}$ data of the central region of M~31 were analyzed for diffuse X-ray emission.
Point sources with the 0.5--10 keV luminosity exceeding $\sim 4 \times 10^{35}$ erg s$^{-1}$ were detected.
Their summed spectra are well reproduced by a combination of a disk black-body component and a black-body component, implying that the emission mainly comes from an assembly of luminous low-mass X-ray binaries.
After excluding these point sources, spectra were accumulated over a circular region of $6\arcmin$ (1.2 kpc) centered on the nucleus.
In the energy range above 2 keV, these residual spectra are understood mainly as contributions of unresolved faint sources and spill-over of photons from the excluded point sources.
There is in addition a hint of a $\sim 6.6$ keV line emission, which can be produced by a hot (temperature several keV) thin-thermal plasma.
Below 2 keV, the spectra involve three additional softer components expressed by thin-thermal plasma emission models, of which the temperatures are $\sim 0.6$, $\sim 0.3$, and $\sim 0.1$ keV.
Their 0.5--10 keV luminosities within 6$\arcmin$ are measured to be $\sim 1.2 \times 10^{38}$ erg s$^{-1}$, $\sim 1.6 \times 10^{38}$ erg s$^{-1}$, and $\sim 4 \times 10^{37}$ erg s$^{-1}$ in the order of decreasing temperature.
The archival ${\it Chandra}$ data of the central region of M~31 yielded consistent results.
By incorporating different annular regions, all the three softer thermal components were confirmed to be significantly extended.
These results are compared with reports from previous studies.
A discussion is presented on the origin of each thermal emission component.

\end{abstract}




\section{INTRODUCTION}

From the era of the ${\it Einstein}$ observatory, it was known through X-ray imaging that there is unresolved and apparently extended X-ray emission in the central region of M~31 \citep{M31_Trinchieri}.
The phenomenon was later confirmed with ${\it ROSAT}$, and it has become generally believed that the emission is contributed not only by faint discrete sources but also by diffuse hot plasmas \citep{M31_Primini,M31_Supper,M31_West,M31_Irwin,M31_diffuse_Borozdin}.

In addition to these soft X-ray imaging studies, detailed wide-band X-ray spectral studies of M~31 have been enabled with the ${\it Ginga}$ satellite; \citet{M31_Makishima} showed that the integrated spectrum of the whole galaxy in the hard 2--20 keV energy band cannot be represented by a power-law or a bremsstrahlung model, but can be reproduced successfully by a physical model developed for Galactic high-luminosity low-mass  X-ray binaries (hereafter LMXBs).
This model consists of a disk black-body (DBB) and a black-body (BB) component, which represent emission from an optically-thick accretion disk and a central neutron star, respectively \citep{LMXB_Mitsuda,LMXB_Makishima,LMXB_Asai}; we hereafter refer to this model as the LMXB model.
This result confirms that bright LMXBs are dominant above the 2 keV energy range, with the 2--20 keV luminosity integrated over the whole M~31 reaching $\sim 5 \times 10^{39}$ erg s$^{-1}$.

Based on these ${\it Ginga}$ results, \citet{Takahashi}, hereafter Paper~1, analyzed the 0.6--10 keV ${\it ASCA}$ spectra integrated over the central 12$'$ (2.4 kpc) radius region of M~31.
This study revealed a clear spectral excess below $\sim$ 1 keV when the ${\it Ginga}$ spectral model mentioned above is extrapolated into the soft energy band.
This excess has been represented successfully with emission from two optically-thin thermal plasmas, of which the temperatures are $\sim$ 0.9 keV and $\sim$ 0.3 keV.
These components are spatially extended at least over the central 2.4 kpc radius, reinforcing the inference from the ${\it Einstein}$ and ${\it ROSAT}$ images that some form of apparently diffuse hot plasma is distributed in the central region of M~31.
The spectral soft excess discovered with ${\it ASCA}$ is also confirmed with ${\it BeppoSAX}$ \citep{M31_diffuse_Trinchieri}.
However, these results have been subject to a possibility that some fraction, if not all, of these thermal soft X-rays are associated with faint discrete sources rather than emitted by truly diffuse large-scale plasmas.

With the advent of the ${\it Chandra}$ and ${\it XMM}$-${\it Newton}$ satellites, it has become possible to study the extended X-ray emission after excluding point sources down to limiting luminosities of $\sim 10^{36}$ erg s$^{-1}$.
\citet[hereafter SEA01]{M31_diffuse_Shirey}
 actually conducted such an analysis on the ${\it XMM}$-${\it Newton}$ spectra integrated over a central 5$\arcmin$ region of M~31, first by excluding bright point sources using the X-ray image, and then modeling the contribution from unresolved faint discrete sources by scaling the average spectrum of the bright sources.
They found only one optically-thin thermal plasma, of which the temperature is $\sim$ 0.35 keV.
This result was apparently reconfirmed by ${\it Chandra}$ \citep{M31_diffuse_Primini,M31_diffuse_Garcia,M31_diffuse_Dosaj}.
The ${\it XMM}$-${\it Newton}$ and ${\it Chandra}$ data thus reveal only one thermal component, while we have found two through the ${\it ASCA}$ spectroscopy (Paper~1).
Then, what causes this difference ?

At first glance, our method employed in Paper~1 may appear much less reliable than the other one, in which all the bright point sources are removed through high-resolution imagery.
Nevertheless, we are concerned about the way employed by SEA01
to estimate the faint source contribution to the residual spectrum.
They estimated this contribution by scaling down a template power-law model, which was determined by fitting a summed spectrum of the removed bright sources.
In reality, this bright-source spectrum is likely to be contaminated by the diffuse component, because of the limited angular resolution of ${\it XMM}$-${\it Newton}$.
Furthermore, a power-law model is known to poorly reproduce the spectra of luminous LMXBs, as pointed out by \citet{LMXB_Makishima} and Paper~1.
For these reasons, we expect that the analyses of the ${\it XMM}$-${\it Newton}$ and ${\it Chandra}$ data can be improved; this has motivated the present paper.

\section{OBSERVATION AND DATA REDUCTION}

In the present paper, we analyze archival data of the M~31 central region obtained by ${\it XMM}$-${\it Newton}$ and ${\it Chandra}$.
Although these datasets have already been analyzed for diffuse emission by other authors
(Garcia et al.\ 2000; Primini et al.\ 2000; SEA01; Dosaj et al.\ 2002),
there still remains room for improved analyses as described in $\S$~1, especially in comparison with Paper~1 and employing more appropriate modeling of the point-source spectra.
Here we describe the observation and data screening.

\subsection{${\it XMM}$-${\it Newton}$ Observation}

\begin{figure}
\begin{center}
\epsscale{0.4}
\plotone{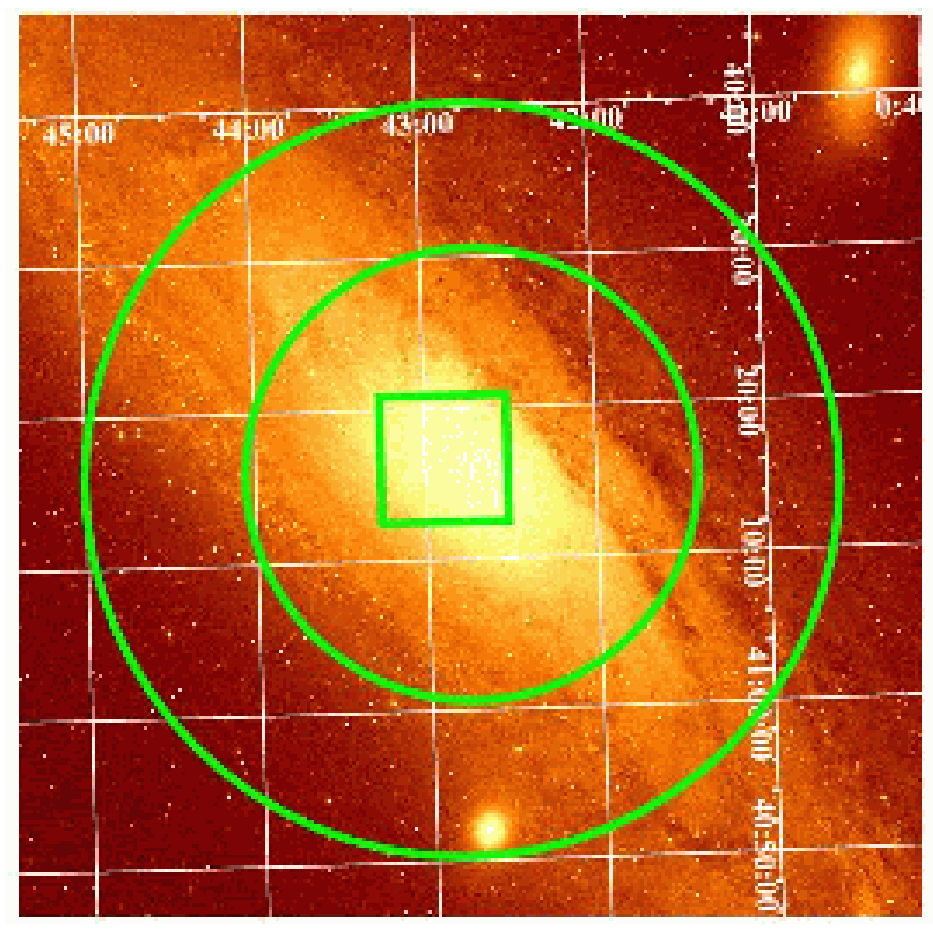}
\end{center}
\caption{X-ray fields of view superposed on a Digitized Sky Survey image of M~31 (${1\arcdeg \times 1\arcdeg}$).
The larger circle represents the field of view of the ${\it ASCA}$/GIS observation (Paper~1), and the smaller one that of the ${\it XMM}$-${\it Newton}$/MOS.
The field of view of the ${\it Chandra}$/ACIS-S3 is represented by the inner square.}
\label{fig:M31_image}
\end{figure}

Among the ${\it XMM}$-${\it Newton}$ archival data, there are four observations of the central region of M~31 (observation ID = 0109270101, 0112570101, 0112570401 and 0112570601), all with the same pointing direction.
While SEA01
 analyzed the 0112570401 dataset as the proprietors, we here analyze the 0112570101 dataset, which has the longest exposure of the four.
The 0112570101 dataset is characterized in Table~\ref{tab:obs}, and its field of view is shown over an optical image in Figure~\ref{fig:M31_image} together with that used for our previous ${\it ASCA}$ results (Paper~1).
The X-ray images obtained by the MOS1, MOS2 and PN detectors are superposed in Figure~\ref{fig:M31_Ximage}a.
As already reported (SEA01),
the image clearly reveals the extended, apparently diffuse X-ray emission in the central region.

We performed SAS standard data screening for the MOS1, MOS2 and PN data.
In this observation, a small background flare occurred only in PN, and the PN events during the flare were discarded; the achieved exposure is 60 ks with each MOS detector, and 49 ks with the PN.
Over the $\sim 15\arcmin$ radius field of view, we detected 212 point sources by the CIAO source detection method (wavdetect) with a detection limit of $\sim 4 \times 10^{35}$ erg s$^{-1}$; point spread functions have no effect on the actual source detection, but they are used in the estimation of the source properties, such as extent or flux and so on, which the wavdetect routine reports (from the CIAO detect manual).
Around each source, we then excluded a circular region of radius $25\arcsec$--$60\arcsec$, as specified by the wavdetect routine referring to the source image size, the source intensity over the surrounding background level, and the off-axis angle.
According to the ${\it XMM}$-${\it Newton}$ users handbook, $\sim$ 85\% of photons from each point source can be excluded by using these radii.
In order to study the extended emission, the remaining events were accumulated over a circular region centered on the nucleus with a radius of 6$\arcmin$.
Then, we constructed energy spectra of individual detectors in the 0.4--7 keV energy band.

While our primary objective is to study the unresolved diffuse emission, it is of similar importance to examine the average spectra of bright point sources.
We have therefore accumulated the events of 92 point sources falling within the 6$\arcmin$ radius, which were once removed in the previous step, and constructed another set of 0.4--10 keV spectra.
To extract the events, we here employed event extraction radii twice as small as those used in deriving the above diffuse spectra; now $\sim$ 60\% of events from each point source are extracted.
These events are dominated by the discrete-source photons, but are contaminated to some extent by the unresolved emission.

To analyze the extended-emission spectra, we generated response matrices (one for each detector) by SAS taking the weighted means of place-by-place responses in reference to the brightness distribution of the extended emission.
To fit the summed point-source spectra, we similarly generated responses as weighted means of those of the individual point sources.
In order to take into account possible uncertainties introduced by these averaging processes, we assign 5\% systematic errors over 0.4--2 keV to both sets of spectra.

Background spectra for both the unresolved and point-source emission were constructed from the corresponding detector regions of blank-sky observation datasets (exposure 350 ks with the PN and 1000 ks with each MOS) provided by \citet{bg_Read}.
In the energy range above 12 keV, the count rate of the derived PN background spectra coincide, within $\sim$ 2\%, with those of the on-source spectra, in agreement with the report by \citet{bg_Katayama} that the EPIC background fluctuates by no more than 8\% (at 1 $\sigma$ level) from observation to observation.
Similarly, the 7--12 keV MOS1 and MOS2 background spectra have been confirmed to agree within 5\% with the extended-emission spectra from the corresponding detectors.
(Such a comparison is difficult for the summed point-source spectra because of the signal detection up to $\sim$ 12 keV.)
Thus, the uncertainties associated with the subtraction of the detector backgrounds are estimated to be significantly smaller than photon counting errors.
We must in addition consider systematic uncertainties in the sky-background subtraction, including in particular the $\sim$ 35\% non-uniformity in the soft X-ray sky brightness \citep{CXB_Lumb}.
However, this is again negligible in the relevant soft X-ray range, where the signals (both extended emission and summed point sources) exceed the background by more than an order of magnitude.

\subsection{${\it Chandra}$ Observation}

There are already many archival ${\it Chandra}$ datasets of the central region of M31, with repeated pointings.
Because we need energy spectra especially in the low energy band, we here select the archived ACIS-S data with the longest exposure (observation ID = 1575; Table~\ref{tab:obs}), which allows us to utilize a larger effective area in energies below $\sim$ 5 keV than the ACIS-I data; in contrast, \citet{M31_diffuse_Garcia}, \citet{M31_diffuse_Primini} and \citet{M31_diffuse_Dosaj} analyzed the ACIS-I data.
In the present analysis, we use only the data from the back-illuminated S3 chip.
Its field of view and X-ray image are presented in Figure~\ref{fig:M31_image} and Figure~\ref{fig:M31_Ximage}b, respectively.

We screened the data in the standard manner, and further discarded the data acquired during small background flares which occurred in the observation.
The usable exposure amounts to 29 ks.
We also removed the events on the columns along the readout streak of the brightest source.
Using the same CIAO wavdetect routine as the ${\it XMM}$-${\it Newton}$ data, 90 point sources were detected over the $8\arcmin \times 8\arcmin$ ACIS-S3 field of view, with the limiting luminosity of $\sim 3 \times 10^{35}$ erg s$^{-1}$ for a typical source spectrum.
This limit is much lower than those of \citet{M31_diffuse_Garcia}, \citet{M31_diffuse_Primini} and \citet{M31_diffuse_Dosaj} who used short (5--10 ks) exposure datasets, and comparable to that of \citet{M31_LF_Kong} who combined 8 separate ACIS-I observations to achieve a total exposure of $\sim$ 40 ks.
We then discarded circular regions of radii $3\arcsec$--$20\arcsec$ (depending on the source image size, the intensity, and the position from the aim point, like in the ${\it XMM}$-${\it Newton}$ case) around the individual sources.
Simulations using the ChaRT software predict that $\sim$ 94\% events from the detected point sources are thus excluded.
Using the remaining events, we constructed an energy spectrum covering a circular region centered on the nucleus within a radius of 3$\arcmin$.
Because the ACIS-S3 field of view is limited to $8\arcmin \times 8\arcmin$ and its aim point is not located at the center of the chip, the full 6$\arcmin$ radius region is unavailable.
Even the 3$\arcmin$ region falls outside the field of view by 6\% at the western edge.
We generated the response matrix by a CIAO weighted averaging method similar to the ${\it XMM}$-${\it Newton}$ case, and applied a correction for the degradation in the low-energy quantum efficiency of the ACIS chips since the launch.

The background spectrum was collected from the corresponding S3 region of blank-sky observation datasets (exposure 450 ks) provided by the ACIS calibration team.
It is known that the background rates of blank-sky data vary, from observation to observation, by up to $\sim$ 40\% in energies below 5 keV.
However, this uncertainty can be ignored in the 0.45--2 keV energy band, because the observed count rate is an order of magnitude higher than the background level.
In the 2--5 keV band where the background systematics are more severe, we confirmed that the blank-sky subtracted spectrum obtained in this way agrees, within 6\%, with that obtained by subtracting another background acquired with the other back-illuminated S1 chip (excluding two point sources) during the same on-source observation; the S1 and S3 backgrounds are generally known to coincide with each other within 3\% in this energy band (http://hea-www.harvard.edu/\~{}maxim/axaf/acisbg/data/README).
Therefore, we here analyze the energy spectrum over the total 0.45--7 keV energy band, after subtracting the blank-sky background spectrum.
Because of relatively poor statistics, we do not utilize these ${\it Chandra}$ data to quantify the spectra of the detected sources.

\section{${\it XMM}$-${\it Newton}$ DATA ANALYSES}

\subsection{Modeling of the Summed Spectra of the Detected Point Sources}

\begin{figure}
\begin{center}
\epsscale{0.4}
\plotone{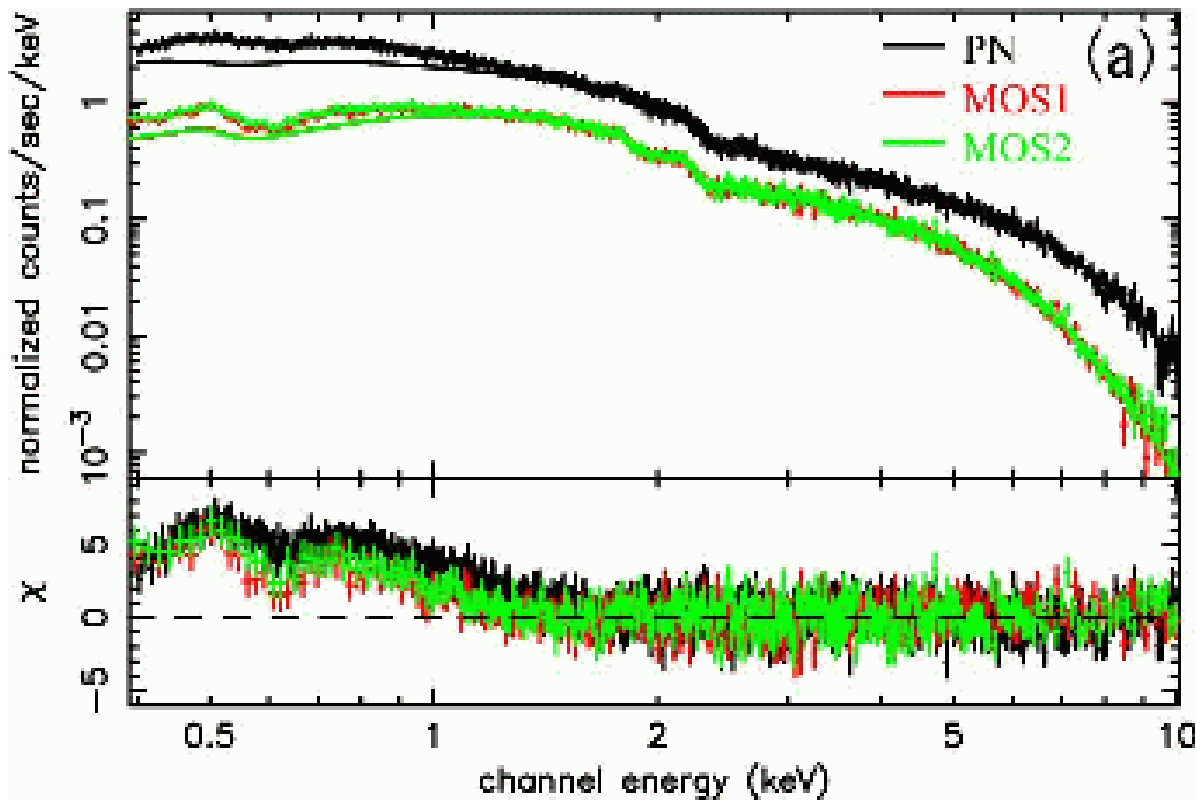}
\plotone{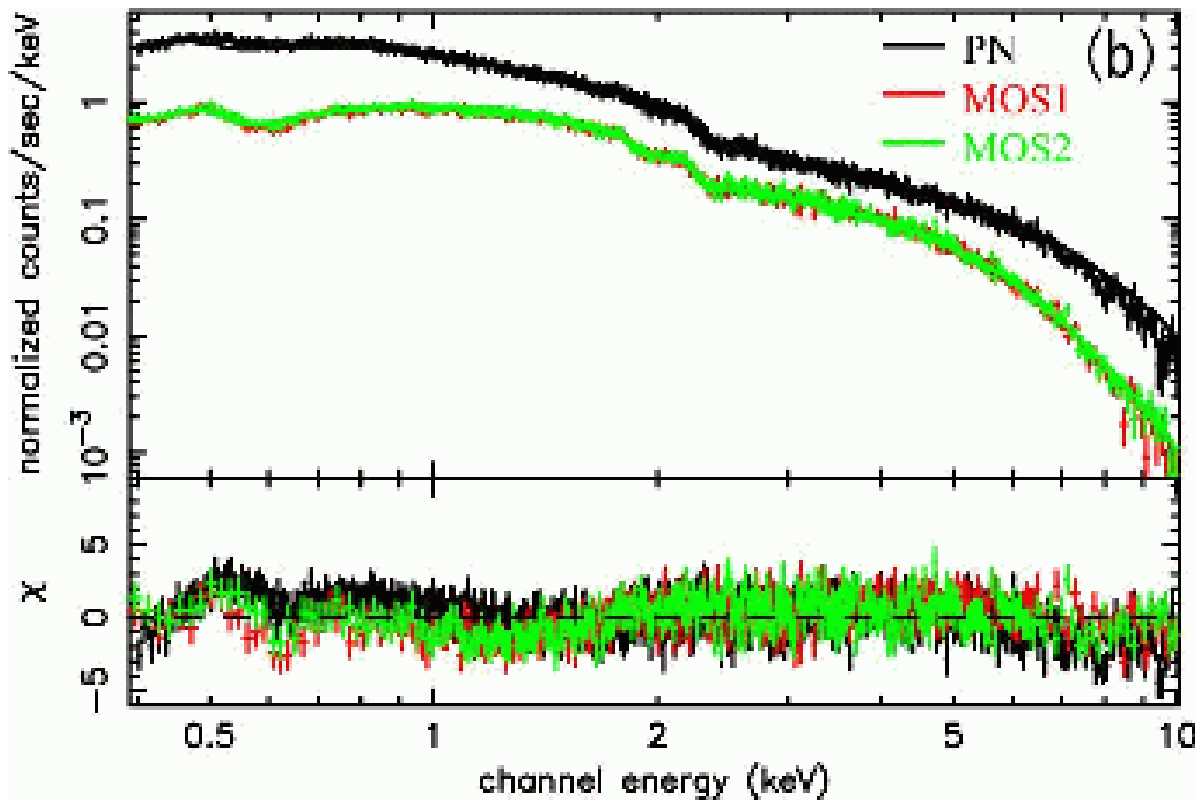}
\plotone{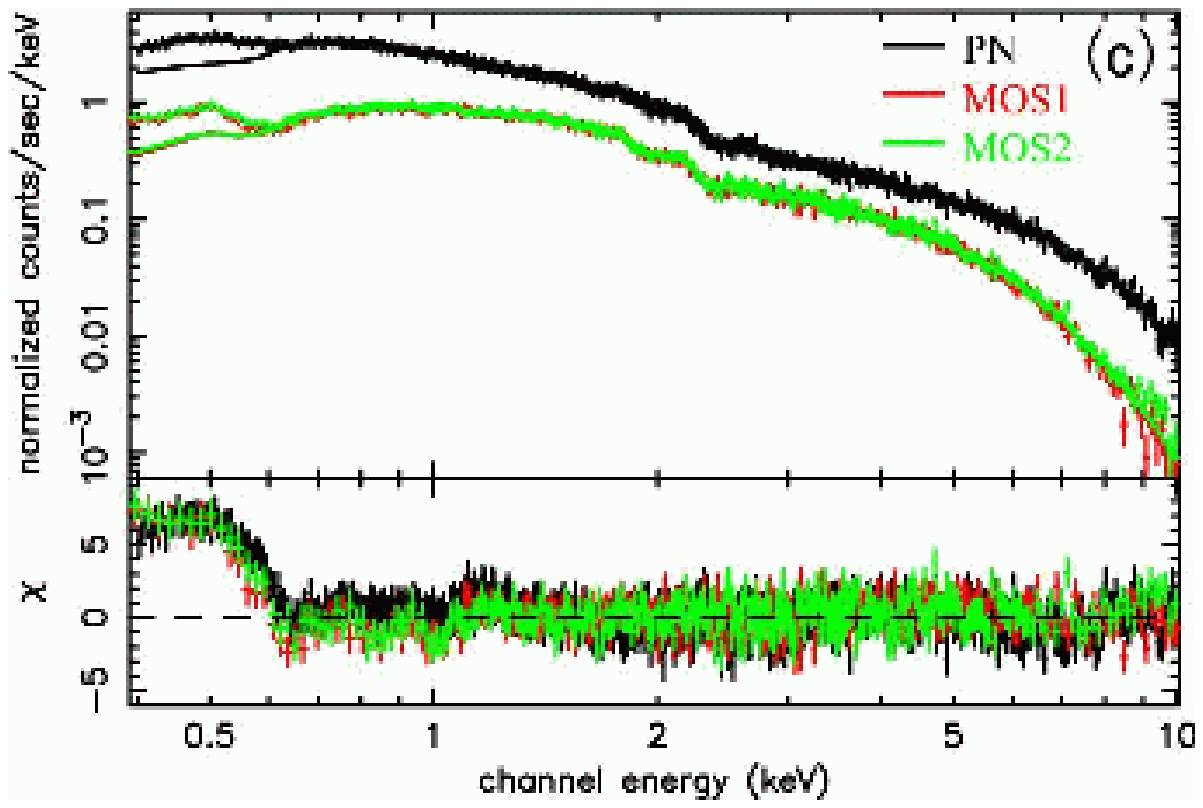}
\end{center}
\caption{${\it XMM}$-${\it Newton}$ PN (black), MOS1 (red) and MOS2 (green) spectra, summed over the 92 point sources detected within the central $6\arcmin$ region of M~31.
They are presented after the background subtraction (see text), but without removing the instrumental response.
They are fitted simultaneously in the energy range above 2 keV with (a) the canonical LMXB model, and above 0.6 keV with (b) the PL and (c) the LMXB+2MKL.
The residuals between the data and the model are shown at bottom of each figure.
The best-fit model is extrapolated to below the fitted energy band to highlight the soft excess, which comes from the diffuse emission ($\S$~3.3).
}
\label{fig:lmxb}
\end{figure}

Although SEA01
 already analyzed the summed spectra of the detected point sources with a single power-law (PL for short) model, we here repeat the analysis for consistency, by considering a small amount of flux contribution from the diffuse X-ray emission.
Figure~\ref{fig:lmxb} shows the MOS and PN spectra obtained in $\S$~2.1, as a sum over the 92 sources detected within $6\arcmin$.
As described in $\S$~2, they have been collected from regions of which the summed area amounts to $\sim$ 10\% of the total $6\arcmin$ radius region.
According to the spectral decomposition in Paper~1, more than half the overall flux of M~31 below $\sim 2$ keV comes from the diffuse component.
From these two facts, the diffuse X-ray emission is expected to contribute $> 5$\% to the spectra of Figure~\ref{fig:lmxb}.
With this in mind, we analyzed the 0.4--10 keV portion of the MOS and PN spectra of Figure~\ref{fig:lmxb}.

We first fitted the spectra above 2 keV, where point sources are dominant (Paper~1), employing the physical LMXB model (a DBB plus a BB) to represent luminous point sources in M~31 ($\S$~1).
We allowed the photoelectric absorption column density to vary freely, but constrained it to be higher than $N_{\rm H} = 6.7 \times10^{20}$ cm$^{-2}$ which corresponds to the Galactic value along the line of sight toward M~31 (from Einline and W3nH).
This LMXB model has successfully reproduced the spectra (Figure~\ref{fig:lmxb}a and Table~\ref{tab:lmxb}), and the obtained temperatures are consistent with those (the innermost disk temperature of $kT_{\rm in} \sim$ 1 keV and the blackbody temperature of $kT_{\rm BB} \sim$ 2 keV) in \citet{M31_Makishima} and Paper~1.
The obtained total luminosity (here and hereafter, we quote luminosities in the 0.5--10 keV range), $1.1 \times 10^{39}$ erg s$^{-1}$, is also consistent with that of the spectral hard component in Paper~1, $1.7 \times 10^{39}$ erg s$^{-1}$; the difference can be explained by a flux leakage ($\sim 40$\%) outside the event accumulation regions around the point sources, and an additional contribution to the ${\it ASCA}$ spectra by sources below the present detection threshold [estimated to be $\lesssim 5 \times 10^{37}$ erg s$^{-1}$ using the Log ${\it N}$--Log ${\it S}$ relation by \citet{M31_LF_Kong}].
Moreover, our luminosity integrated over the $6\arcmin$ radius compares reasonably with the 2--20 keV value of the whole M~31 derived with ${\it Ginga}$, $5 \times 10^{39}$ erg s$^{-1}$ \citep{M31_Makishima}.
Following these consistent results among the three satellites, we hereafter fix $kT_{\rm BB}$ at 2 keV, to prevent the value from becoming unconstrained when the soft energy band with much higher statistics is included.
When we re-fitted the 2--10 keV MOS and PN spectra by replacing the LMXB model with a PL (Table~\ref{tab:lmxb}), the model was too flat to reproduce the spectra in the hard energy band as mentioned in Paper~1.

When the successful LMXB fit determined in the 2--10 keV band is extrapolated down to the same lower energy bound of 0.6 keV as employed in the ${\it ASCA}$ analyses, the data exhibit excess above the model prediction (Figure~\ref{fig:lmxb}a).
If we re-adjust the model parameters except $kT_{\rm BB}$ fixed at 2 keV, the fit is improved but still remains unacceptable (Table~\ref{tab:lmxb}).
Restoring the PL model, which was once rejected, gives consistent parameters as obtained by SEA01 [the photon index of $1.82\pm0.03$ and the absorption column density $(6.7\pm0.4)\times 10^{20}$ cm$^{-2}$], but the fit is again unacceptable (Figure~\ref{fig:lmxb}b and Table~\ref{tab:lmxb}).
Moreover, essentially the same problem is also seen in the discrete-source spectra analyzed by SEA01; the data deviate from the model at 1 keV, like in our Figure~\ref{fig:lmxb}b.

Since we expect that the above fit failures are caused by a small amount of diffuse flux which leaked into the regions around the point sources, we have added two thin-thermal plasma emission components to the LMXB fit after the ${\it ASCA}$ results (Paper~1).
As the plasma emission code, we use the MEKAL model
\citep[MKL for short;][]{MEKALI_Mewe,MEKALII_Mewe,MEKAL_Kaastra,MEKAL_Liedahl}
with the improved Fe-L modeling, instead of the Raymond-Smith code
\citep[RS for short;][]{RS_Raymond}
employed in Paper~1.
The absorption within M~31 is thought to be rather low, so we here and hereafter fix the absorption of thin-thermal plasmas at the Galactic value [$(6.7\pm0.4)\times 10^{20}$ cm$^{-2}$].
For the moment, we constrain for simplicity the plasma abundances at 1 solar in addition.
This model, denoted LMXB+2MKL model, has successfully reproduced the spectra above 0.6 keV (Figure~\ref{fig:lmxb}c), and the measured LMXB and two MKL temperatures are consistent with those in Paper~1 (Table~\ref{tab:lmxb}).
Thus, the LMXB model, when softer MKL contributions are included, can successfully reproduce the entire 0.6--10 keV spectra of the summed point sources.

Although thus successful in energies above 0.6 keV, the LMXB+2MKL model again falls below the observed spectra, when extrapolated toward lower energies.
This can be attributed to yet another diffuse component of a still lower temperature leaking into the spectra, as identified later in $\S$~3.3 using the diffuse-emission spectra.

\subsection{Modeling of the 0.6--7 keV Spectra of the Diffuse X-ray Emission}

\begin{figure}
\begin{center}
\epsscale{0.4}
\plotone{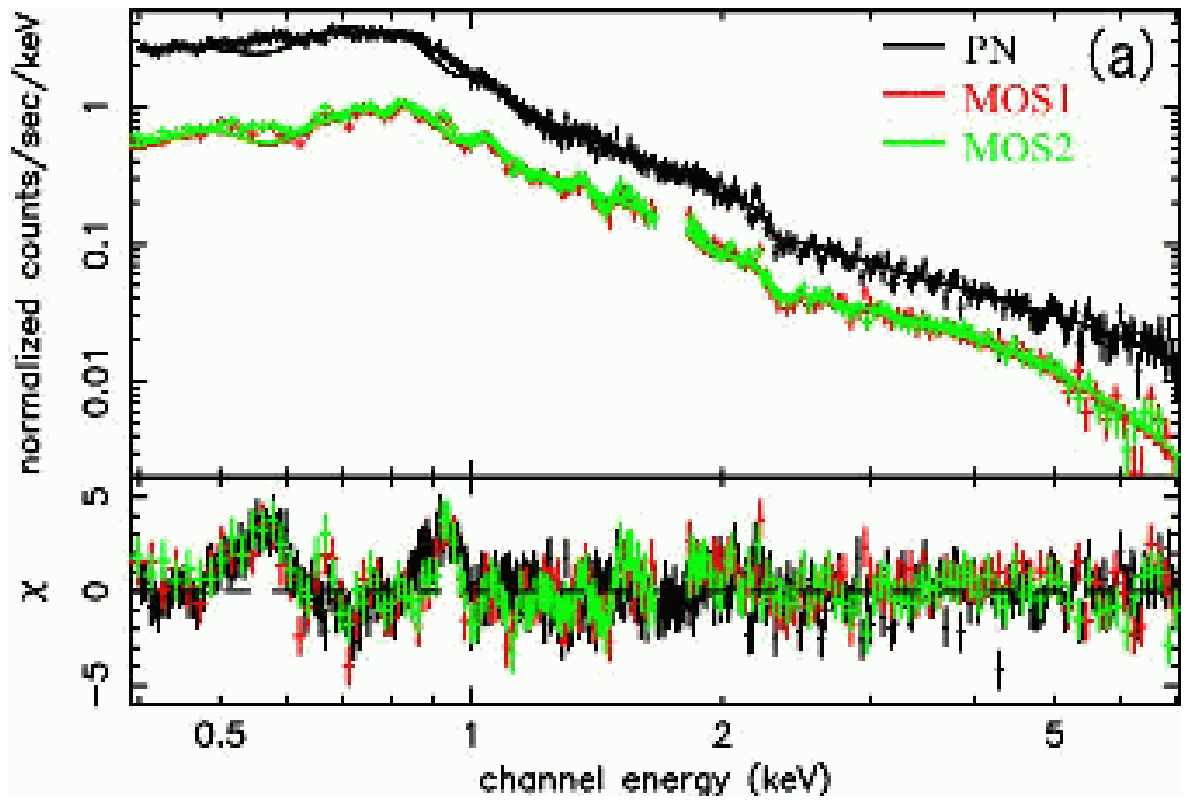}
\plotone{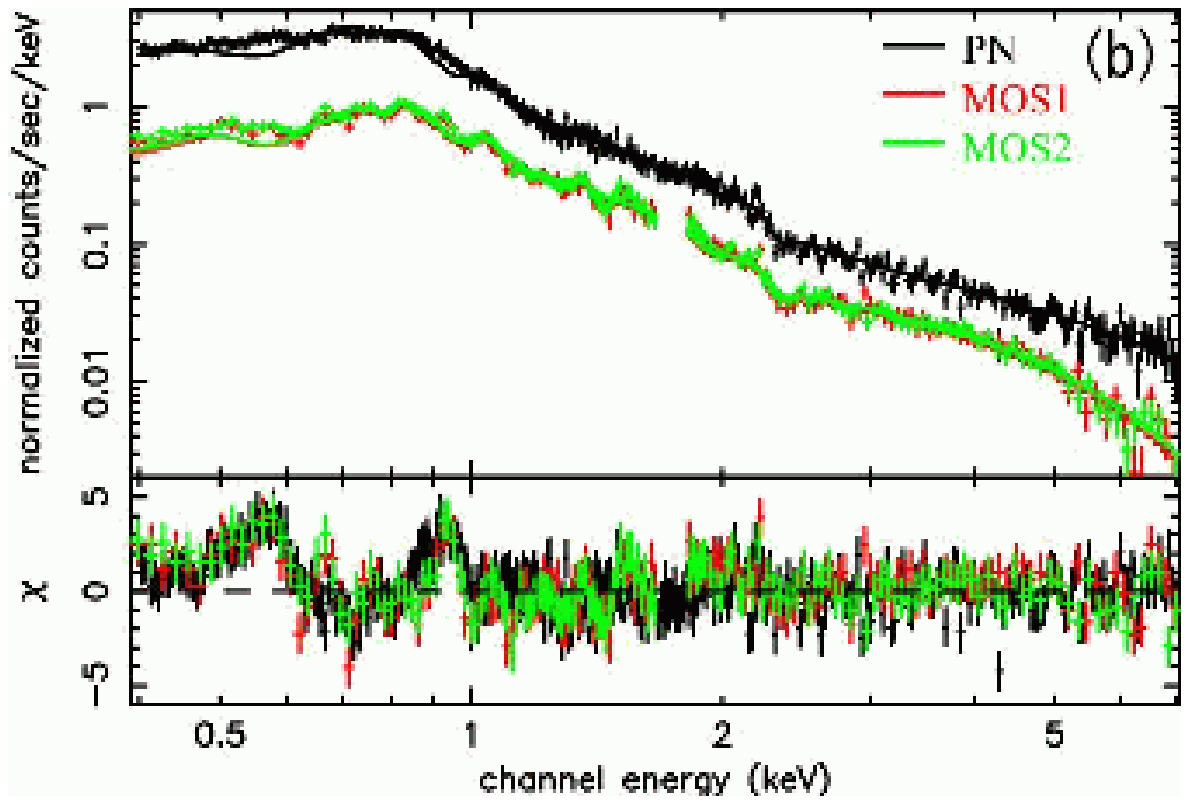}
\plotone{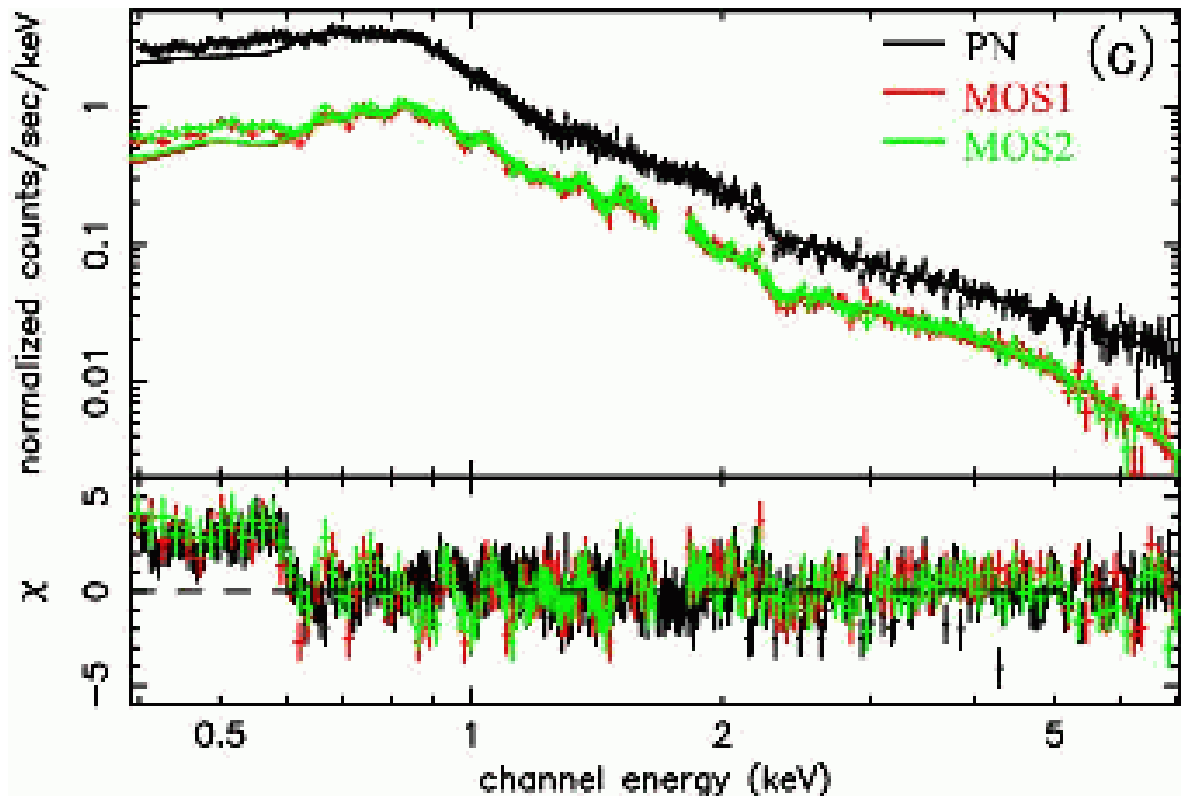}
\plotone{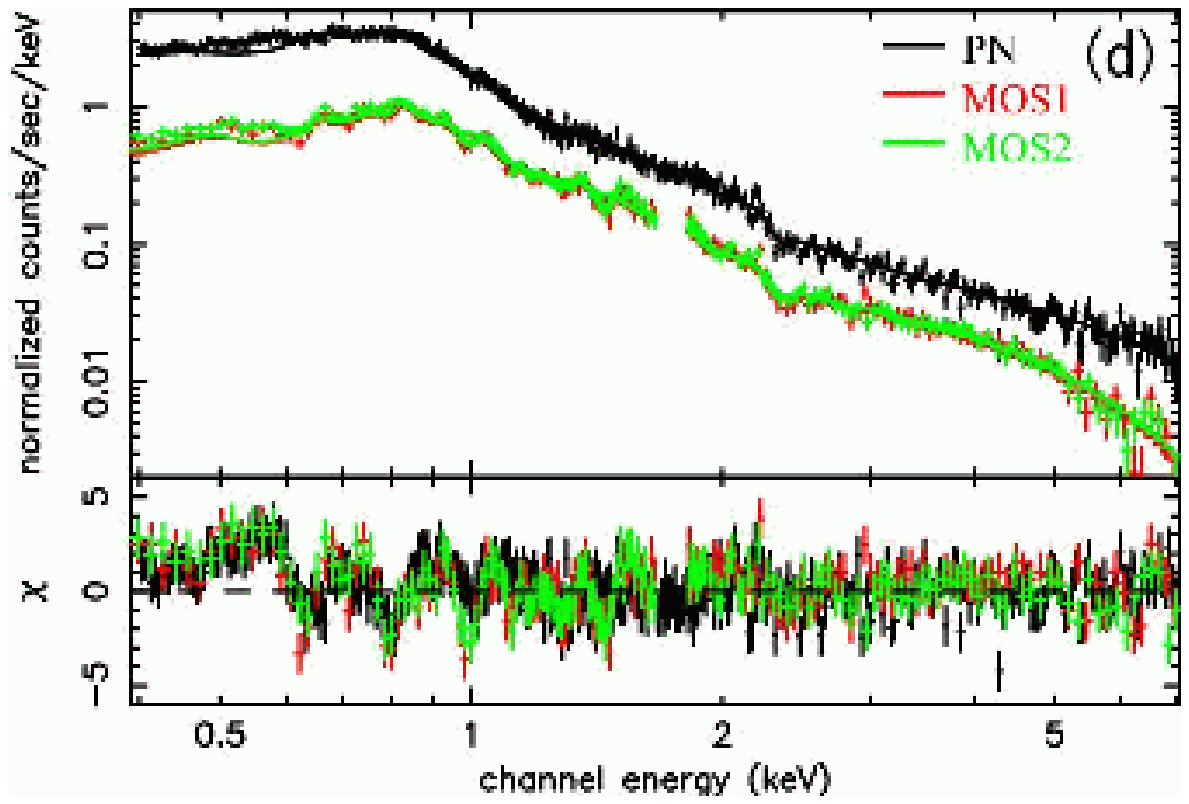}
\end{center}
\caption{Background-subtracted ${\it XMM}$-${\it Newton}$ PN (black), MOS1 (red) and MOS2 (green) spectra of the central 6$\arcmin$ region of M~31, obtained by removing the detected point sources.
The instrumental response is included.
They are fitted simultaneously in the energy range above 0.6 keV with; (a) the LMXB+1MKL, (b) the PL+1MKL, (c) the LMXB+2MKL, and (d) the LMXB+2RS model.
The best-fit models are extrapolated to below 0.6 keV to highlight the soft excess.
}
\label{fig:0.6-7keV}
\end{figure}

In order to quantify the diffuse X-ray emission using the ${\it XMM}$-${\it Newton}$ data, we jointly analyzed the residual MOS and PN spectra of the central 6$\arcmin$ region, obtained by removing the point sources as described in $\S$~2.
The spectra, shown in Figure~\ref{fig:0.6-7keV}, are considerably softer than those for bright point sources (Figure~\ref{fig:lmxb}), and bear a clear hump over 0.7--1.0 keV due to the Fe-L complex indicative of the dominance of soft thermal emission.
Moreover, the spectra extend significantly toward higher energies than would be expected for the softer thermal emission.
This is mainly considered to have two origins.
One is the $\sim 15\%$ spill-over of photons from the detected bright point sources out of the accumulation regions; this is estimated to contribute $\sim 2.6 \times 10^{38}$ erg s$^{-1}$.
The other is contribution from discrete sources below our detection limit, which is estimated to be $\lesssim 5 \times 10^{37}$ erg s$^{-1}$ based on the Log ${\it N}$--Log ${\it S}$ relation of \citet{M31_LF_Kong}.
Since the former dominates, we model the hard component in the ``diffuse'' spectra (Figure~\ref{fig:0.6-7keV}) by the LMXB model in contrast to SEA01 who used a PL modeling, which is ruled out in our analyses ($\S$~3.1).
Because the absorption column density of the summed point-source spectra is consistent with the Galactic value (Table~\ref{tab:lmxb}), we here and hereafter apply the Galactic $N_{\rm H}$ to the LMXB component, as well as to the thin-thermal plasma components ($\S$~3.1).
The abundances of elements are allowed to vary freely, but their relative values are constrained to follow the solar ratios.

We tentatively limited the fit energy band to 0.6--7 keV, to compare with the ${\it ASCA}$ results (Paper~1).
The Si-K energy band (1.68--1.80 keV) of both MOS detectors was excluded, because the background contribution and uncertainty are rather large there \citep{CAL_Ferrando}.
Figure~\ref{fig:0.6-7keV}a shows the simplest model, i.e., one LMXB plus one-temperature MKL model (LMXB+1MKL model), fitted simultaneously to the MOS and PN spectra over the 0.6--7 keV range.
The obtained parameters are summarized in Table~\ref{tab:fit}.
Thus, an MKL component with the temperature 0.3--0.4 keV roughly accounts for the softer part of the spectrum, but there remains excess around 0.9 keV, which makes the fit statistically unacceptable.
If we use the same model as SEA01, namely one PL plus one MKL model (PL+1MKL),
the PL photon index ($\sim 1.8$) and the MKL temperature ($\sim 0.4$ keV) become close to their results (Figure~\ref{fig:0.6-7keV}b and Table~\ref{tab:fit}), but the excess remains as well.
Since this energy band is dominated by the Fe-L line complex which is rather sensitive to the temperature, the failure of the LMXB+1MKL (or PL+1MKL) model suggests, as we expected, that the thin thermal plasma distributed in the central $\sim$ 1.2 kpc region of M~31 cannot be described by a single temperature.

To improve the model, we added another MKL component.
This model, LMXB+2MKL model, is essentially the same as that used in Paper~1, except the plasma codes.
As presented in Figure~\ref{fig:0.6-7keV}c and Table~\ref{tab:fit}, this model has successfully reproduced the 0.6--7 keV spectra, yielding two temperatures of $\sim 0.6$ keV and $\sim 0.3$ keV.
These results reconfirm the conclusion of Paper~1 that two sub-keV temperatures are needed to reproduce the 0.6--10 keV ${\it ASCA}$ over a 12$\arcmin$ radius region around the M~31 nucleus.

Although the temperature of the hotter plasma ($\sim 0.6$ keV) obtained in this way is a little lower than that in Paper~1 (0.9 keV), the disagreement can be attributed to the difference of the plasma models.
Actually, it becomes 0.8 keV (Figure~\ref{fig:0.6-7keV}d and Table~\ref{tab:fit}), in good agreement with Paper~1, when we restore the RS models (LMXB+2RS model) instead of the MKL (LMXB+2MKL).
We tentatively identify the present 0.6 keV component with the 0.9 keV one found with ${\it ASCA}$.
Incidentally, the RS model gives a worse $\chi^{2}$, due to a residual structure around 1 keV which probably arises from an insufficient modeling of the Fe-L complex.

\subsection{Modeling of the 0.4--7 keV Spectra of the Diffuse X-ray Emission}

When the LMXB+2MKL model determined in the 0.6--7 keV band is extrapolated toward lower energies, the model prediction falls significantly short of the actual data (Figure~\ref{fig:0.6-7keV}c).
As a result, the LMXB+2MKL fit actually becomes unacceptable when the 0.4--0.6 keV band is included.
The fit is improved by re-adjusting the model parameters except $N_{\rm H}$ and $kT_{\rm BB}$.
Nevertheless, the temperature of the cooler MKL component becomes lower than that obtained in $\S$~3.2, and the abundances are too low.
When the temperature and abundances are fixed even at 0.25 keV and the 0.1 solar values, respectively, the fit becomes unacceptable ($\chi^{2}$/d.o.f.\ = 819/706).
In short, the LMXB+2MKL modeling does not give a reasonable account of the 0.4--7 keV spectra.
This in turn suggests the presence of a forth, and the softest, emission component, which was not detected with ${\it ASCA}$ (Paper~1), because of the limited soft X-ray sensitivity.

\begin{figure}
\begin{center}
\epsscale{0.4}
\plotone{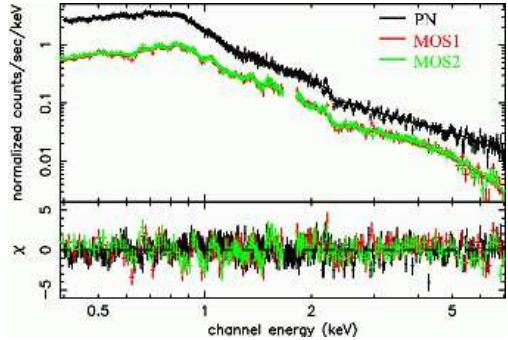}
\end{center}
\caption{Same as Figure~\ref{fig:0.6-7keV}, but the fit is performed over the full 0.4--7 keV energy range using the LMXB+3MKL model.}
\label{fig:0.4-7keV}
\end{figure}

The softest component may be contributed by O-K lines appearing in the 0.5--0.6 keV range.
We therefore added one more MKL component to jointly fit the total-band (0.4--7 keV) MOS/PN spectra.
Here and hereafter, we allow the nitrogen abundance to vary freely, without obeying the solar ratio, because its K-lines appear at the softest end of the spectra, where the instrumental uncertainty is rather large \citep{CXB_Lumb} and yet the data statistics are very high.
We do not address the obtained nitrogen abundance, because it does not deviate much from 1 solar.
This model, denoted as LMXB+3MKL model, has indeed decreased $\chi^2$, as shown in Figure~\ref{fig:0.4-7keV} and Table~\ref{tab:fit}, and has made the fit acceptable.
According to an ${\it F}$-test, the $\chi^2$ difference is significant at a $>$ 99\% confidence level.
The obtained parameters are also consistent with those in $\S$~3.2.
Thus, the diffuse X-ray emission from the central region of M~31 can be understood as a sum of the three components with the temperatures of $\sim 0.6$, $\sim 0.3$, and $\sim 0.1$ keV.
Below, we assess the credibility of this interpretation from several aspects.

\begin{figure}
\begin{center}
\epsscale{0.4}
\plotone{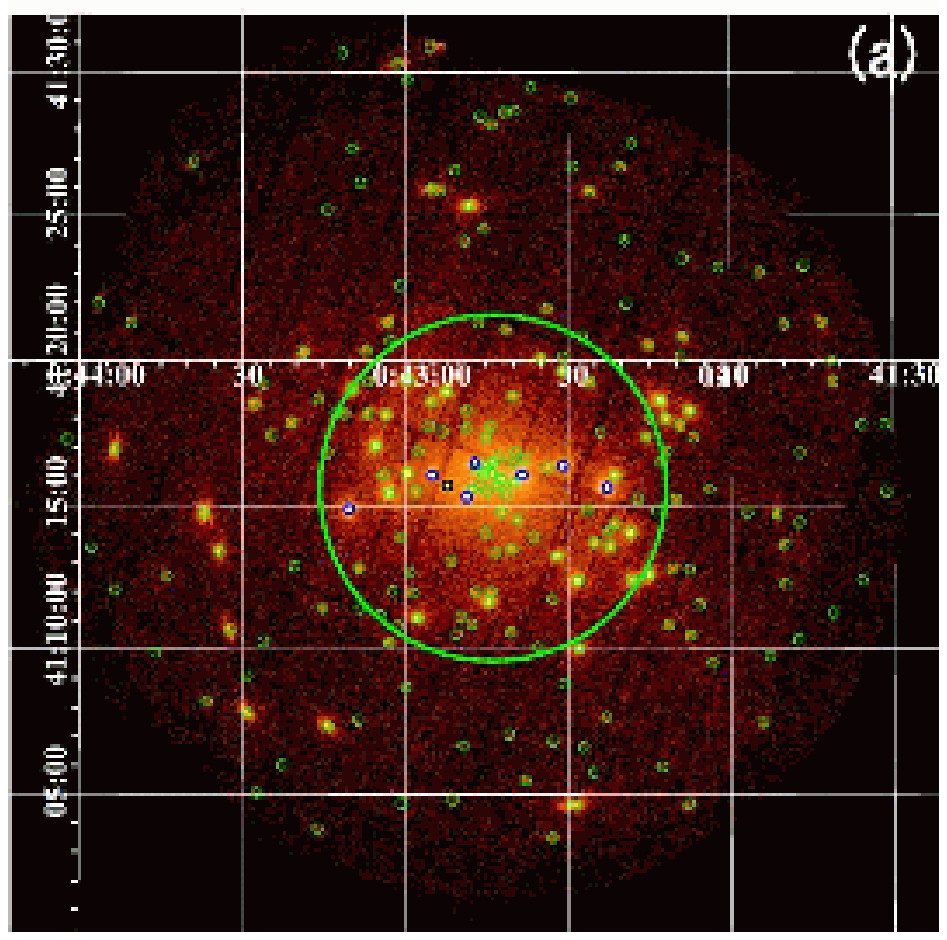}
\plotone{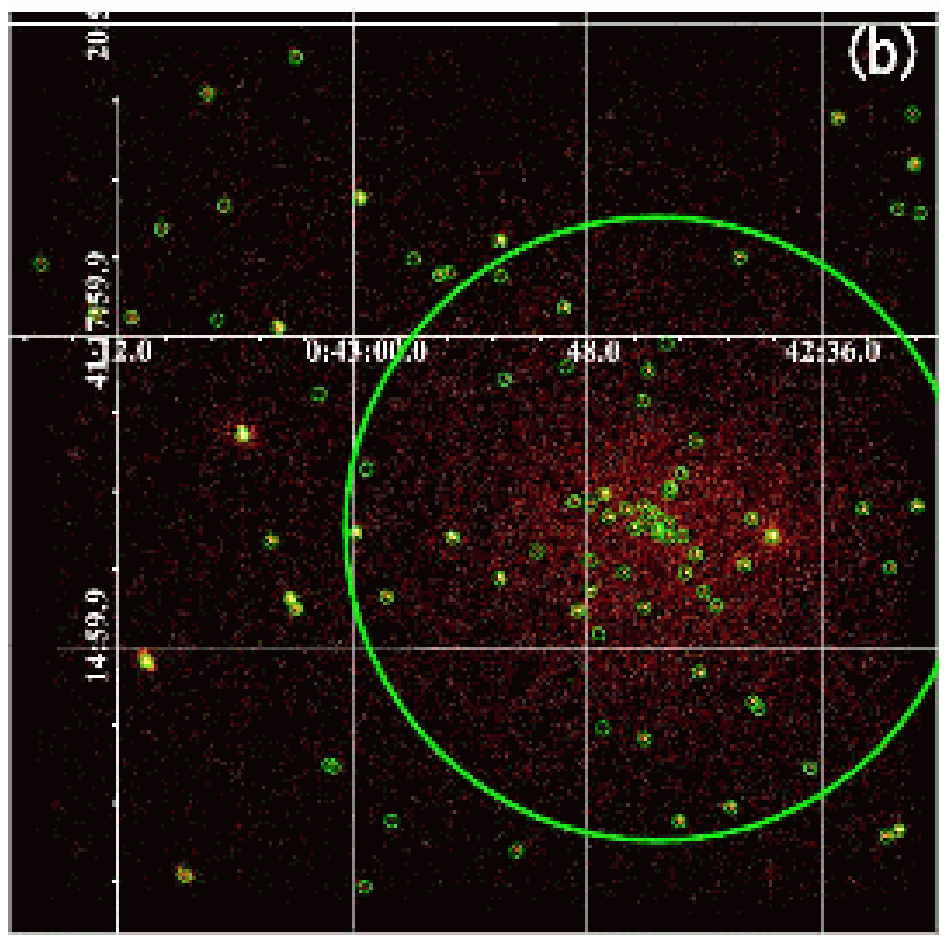}
\end{center}
\caption{X-ray images of M~31 obtained within the fields of view of (a) the ${\it XMM}$-${\it Newton}$/EPIC (the PN and two MOS images are superposed) and (b) the ${\it Chandra}$/ACIS-S3, in the energy ranges of 0.4--7 keV and 0.45--7 keV, respectively.
The background is included.
The sky coordinates are J2000.
In each image, the large circle represents the 6$\arcmin$/3$\arcmin$ (for ${\it XMM}$-${\it Newton}$/${\it Chandra}$) radius region centered on the nucleus, and the other small circles represent the point sources which are excluded.
The 7 blue and 1 black circles indicate the 7 brightest sources within 6$\arcmin$ and the SSS respectively, analyzed in $\S$~3.3.
}
\label{fig:M31_Ximage}
\end{figure}

We have so far fixed the absorption column density of the LMXB model at the Galactic value.
However, some point sources may suffer excess absorption within M~31.
In order to examine such a possibility, we analyzed the 0.4--10 keV spectra of the 7 brightest point sources within the 6$\arcmin$ radius regions (blue circled in Figure~\ref{fig:M31_Ximage}a), of which the summed flux accounts for $\sim$ 50\% of that analyzed here.
The LMXB model successfully reproduced their spectra in general, yielding the absorption column densities in the range (1--6) $\times10^{20}$ cm$^{-2}$ with a typical uncertainty of $1 \times10^{20}$ cm$^{-2}$.
As another examination, we tentatively constrained the LMXB absorption at the three times larger or smaller values ($20 \times10^{20}$ cm$^{-2}$ or $2 \times10^{20}$ cm$^{-2}$) while retaining those of the MKL components at the Galactic value.
However, the LMXB+3MKL fits did not change significantly ($\chi^{2}$/d.o.f.\ = 772/702 and 769/702, respectively), and the overall model parameters remained unchanged within $\lesssim$ 10\%.
This result is easily understandable, because the observed diffuse signal below $\sim$ 1 keV is comparable to that from the summed point sources ($\S$~3.1), and hence is $\sim$ 4 times higher than that estimated from the spill-over of the excluded point sources.
In short, we are thus justified in fixing the absorption of the LMXB component at the Galactic value.

The derived MKL abundances are rather low at their face values.
However, fixing them at 0.3 solar worsen the fit only slightly, to $\chi^{2}$/d.o.f.\ = 792/703, without significant changes in the other parameters.
Considering the complexity of our final spectral model, we hence put reservations on the metallicity of our thermal components.

At $\sim 1\arcmin.5$ south-east of the nucleus (indicated with a black circle in Figure~\ref{fig:M31_Ximage}a), a super soft source (SSS) has been detected \citep{M31_LF_Kong}.
In our data, it is luminous (absorbed luminosity $\sim 2 \times 10^{37}$ erg s$^{-1}$ in 0.3--2 keV), and its spectrum can be reproduced by the same modeling as that used by \citet{M31_SSS_Stefano}, namely the sum of a BB (temperature $\sim 53$ eV) and a PL (photon index fixed at 2.0) with the absorption column density $\sim 4.3 \times 10^{21}$ cm$^{-2}$.
Although we excluded this SSS, the $\sim 15\%$ spill-over of its counts may be contributing to the softest component, because the spectral shape of the SSS is totally different from that of the bright LMXBs and its contribution cannot be accounted for by the LMXB model.
Accordingly, we repeated the LMXB+3MKL fit, but further adding an absorbed ($N_{\rm H}$ = $4.3 \times 10^{21}$ cm$^{-2}$) BB of which the temperature is fixed at 53 eV and the normalization is constrained at 15\% of the SSS spectrum.
The PL component of the SSS is negligible.
As a result, the luminosity of each MKL component has decreased by only 2--4$\%$, and the other parameters have remained unchanged within the errors.
Therefore, we ignore the SSS spill-over.

\section{${\it Chandra}$ DATA ANALYSES}

\begin{figure}
\begin{center}
\epsscale{0.4}
\plotone{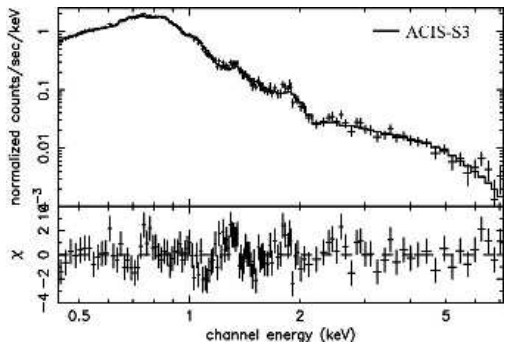}
\end{center}
\caption{Background-subtracted ${\it Chandra}$ ACIS-S3 spectrum of the central 3$\arcmin$ radius region of M~31, obtained by removing the detected point sources.
The instrumental response is included.
It is fitted over the full 0.45--7 keV energy range using the LMXB+3MKL model.}
\label{fig:0.45-7keV}
\end{figure}

In order to confirm our ${\it XMM}$-${\it Newton}$ results under a more stringent point-source elimination, we have analyzed the 0.45--7 keV ${\it Chandra}$ spectrum of the central 3$\arcmin$ region, obtained after removing the point sources as described in $\S$~2.2.
The total area removed by this process is similar between ${\it Chandra}$ and ${\it XMM}$-${\it Newton}$, because the larger areas of the ${\it XMM}$-${\it Newton}$ data around individual sources are approximately compensated by the smaller number of detected point sources in the crowded region around the nucleus.
The derived ACIS spectrum is shown in Figure~\ref{fig:0.45-7keV}.
The spectral fitting was executed in the same way as those of ${\it XMM}$-${\it Newton}$, namely the LMXB plus a few MKL models with the Galactic absorption, because the $\sim$ 6\% spill-over of the excluded point sources is still dominant over the sum of undetected faint sources.
The former is estimated to be $\sim 4.5 \times 10^{37}$ erg s$^{-1}$, by multiplying the 0.5--10 keV total luminosity of the detected point sources ($7.5 \times 10^{38}$ erg s$^{-1}$; converted from the observed count rates output by the wavdetect routine in $\S$~2.2), with the spill-over ratio of the ${\it Chandra}$ X-ray mirror ($\sim 6\%$), while the latter is estimated as $\lesssim 1 \times 10^{37}$ erg s$^{-1}$, by extrapolating the luminosity function obtained by \citet{M31_LF_Kong}.

Table~\ref{tab:fit2} lists the results of fitting the 0.45--7 keV ACIS-S spectrum with four trial models.
While the single LMXB model is far from being acceptable, the addition of one MKL component drastically improve the fit, reconfirming the dominance of thermal X-rays in this ``diffuse'' spectrum.
The $\chi^{2}$ decreases introduced by the 2nd and 3rd MKL components are also significant at $>$ 99 \% confidence levels according to the ${\it F}$-test.
The obtained temperatures of the three MKL plasmas are $\sim 0.6$, $\sim 0.3$, and $\sim 0.1$ keV with the $\sim$ 0.3 solar abundance,
 in good agreement with the ${\it XMM}$-${\it Newton}$ results ($\S$~3.3).
Thus, it is confirmed that the ${\it Chandra}$ spectrum within the central 3$\arcmin$ ($\sim$ 0.6 kpc) region also contains the three thin-thermal plasma components.

\section{SPATIAL DISTRIBUTIONS OF INDIVIDUAL PLASMA COMPONENTS}

\begin{figure}
\begin{center}
\epsscale{0.4}
\plotone{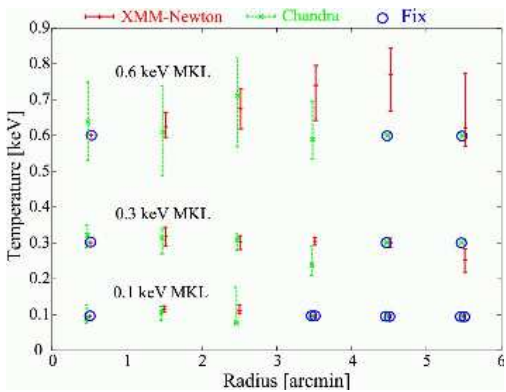}
\end{center}
\caption{The three MKL temperatures of 6 ${\it XMM}$-${\it Newton}$/${\it Chandra}$ annular spectra obtained with the LMXB+3MKL model, presented as a function of the radius from the nucleus.
Small circles indicate that these temperatures are fixed at 0.6, 0.3 or 0.1 keV, when executing the spectral fittings.
}
\label{fig:temperature}
\end{figure}

To confirm the reality and the nature of the four spectroscopic components constituting the LMXB+3MKL model, we next study their radial brightness distributions.
Accordingly, we again produced the ${\it XMM}$-${\it Newton}$ spectra (together with the blank-sky backgrounds and response matrices) of the extended emission in the same way as in $\S$~2.1, but using 6 annular regions of 1$\arcmin$ width each, from the nucleus to a radius of 6$\arcmin$.
These spectra have actually been well ($\chi^{2}$/d.o.f.\ = 1.09--1.16) reproduced by the same LMXB+3MKL model under the same conditions as used in $\S$~3.4 and 4, except that the metal abundances other than nitrogen are fixed at 0.3 solar units.
Similarly, the ${\it Chandra}$/ACIS-S3 spectra derived from the same annular regions were fitted successfully ($\chi^{2}$/d.o.f.\ = 0.99--1.26) by the same LMXB+3MKL model.
In these fits to the EPIC and ACIS spectra using the different annuli, the three MKL temperatures have always been found at $\sim$ 0.6, $\sim$ 0.3 and $\sim$ 0.1 keV as shown in Figure~\ref{fig:temperature}, although we sometimes had to fix some MKL temperatures at their canonical values because of poor statistics.

\begin{figure}
\begin{center}
\epsscale{0.4}
\plotone{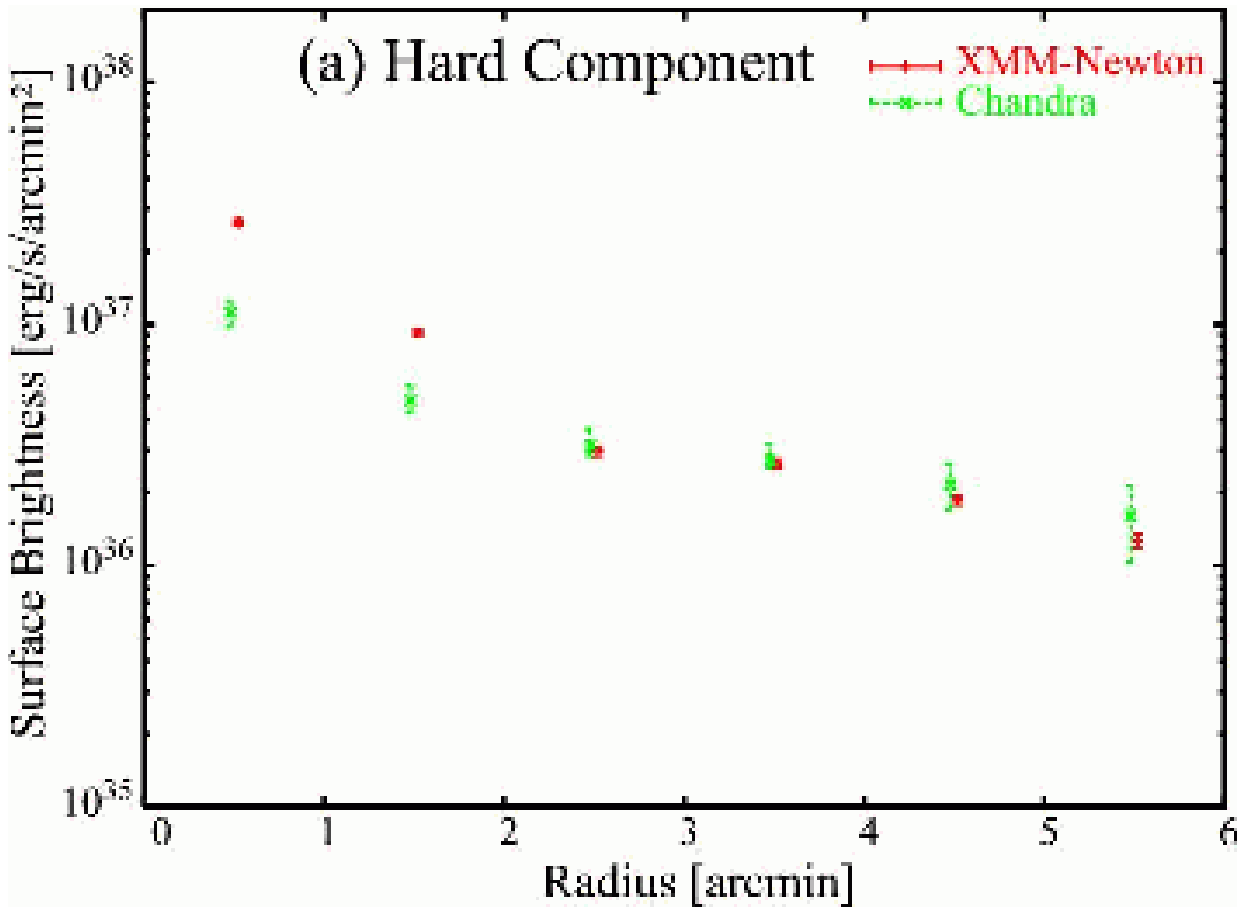}
\plotone{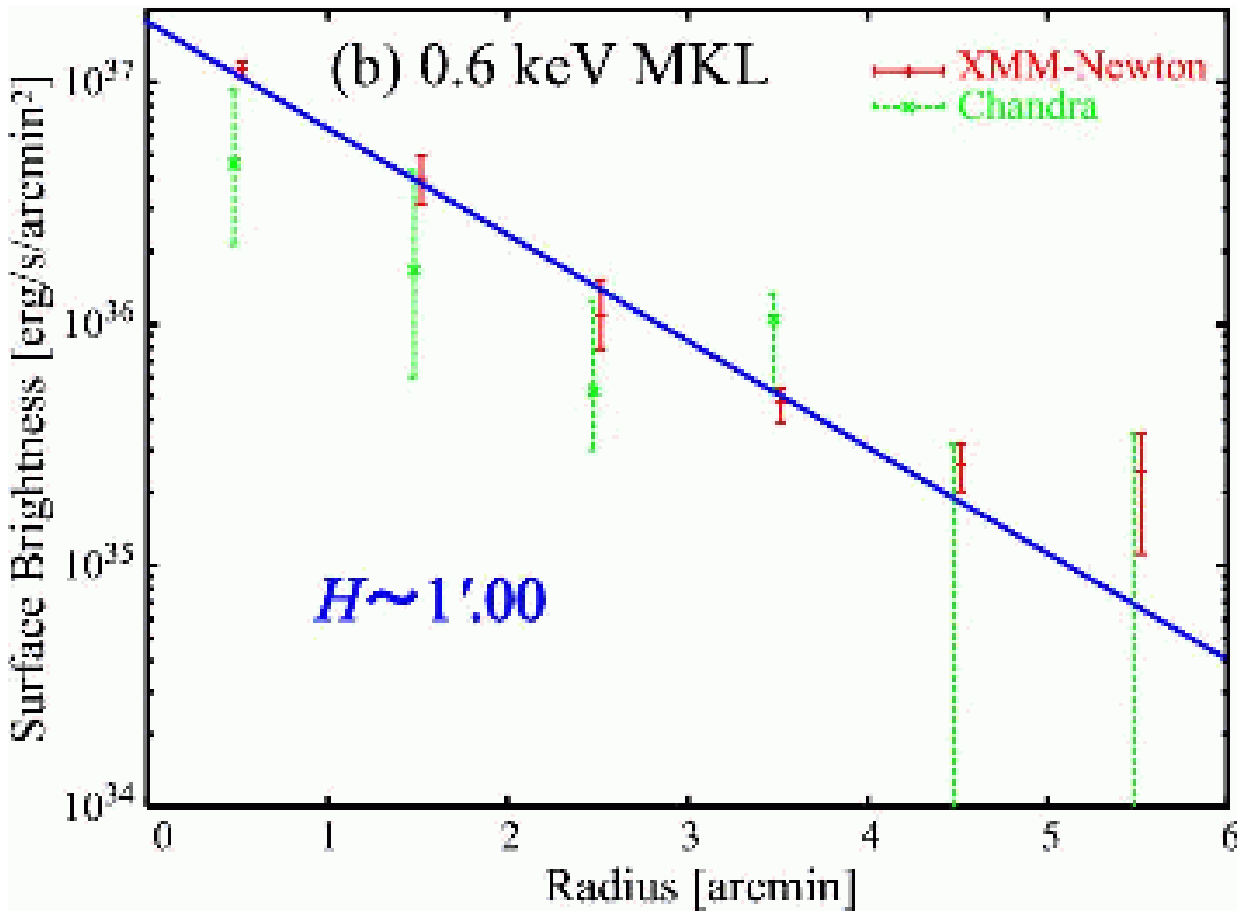}
\plotone{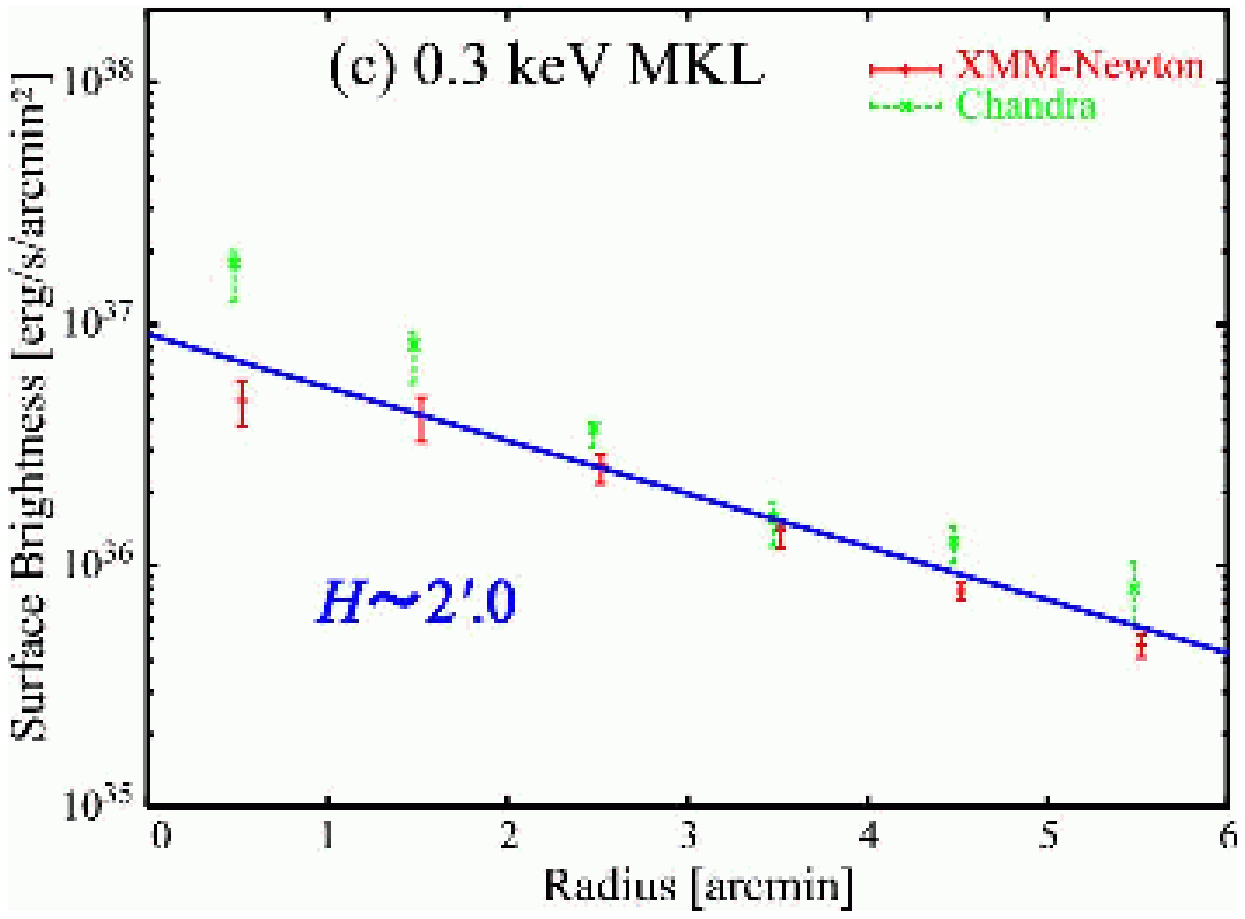}
\plotone{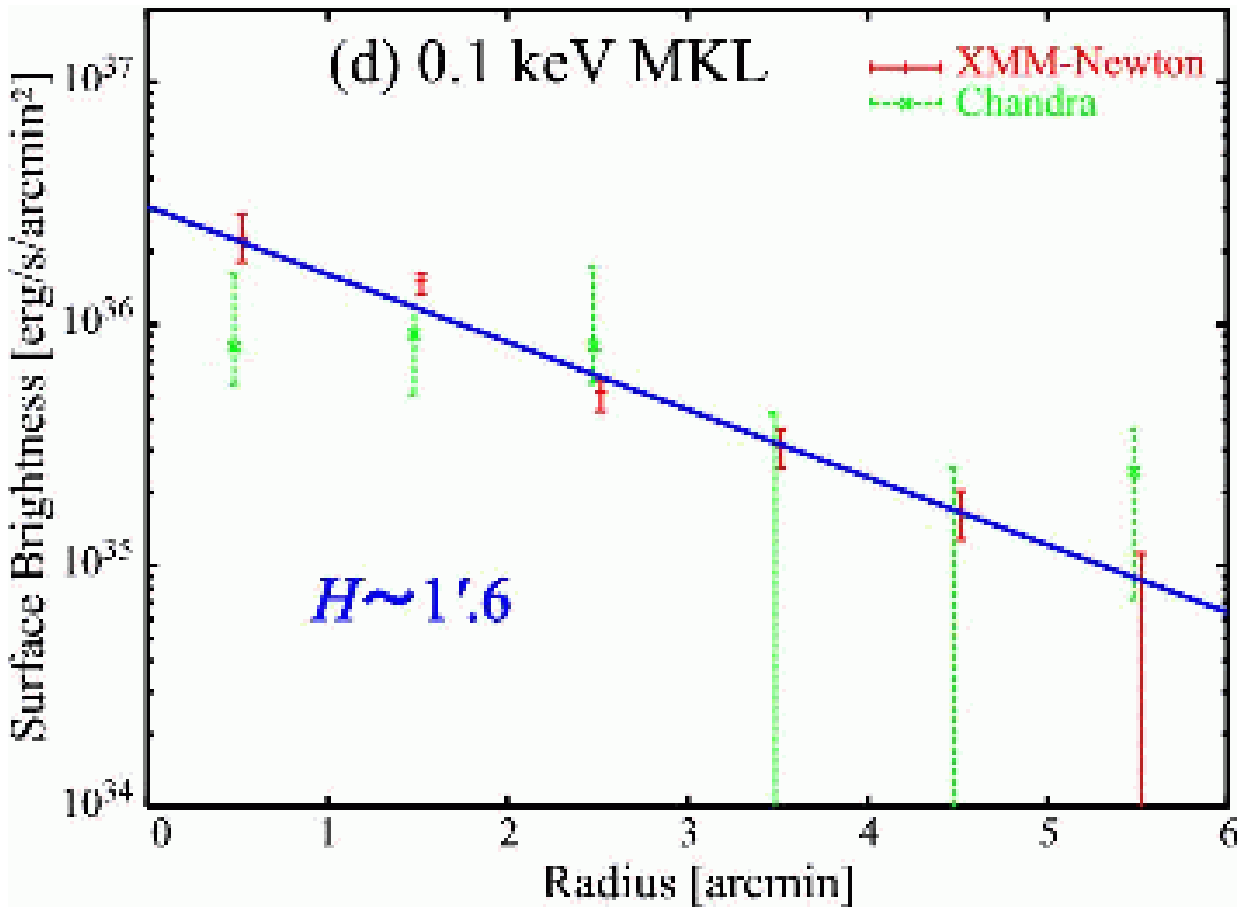}
\end{center}
\caption{The 0.5--10 keV surface brightness of the four X-ray emission components of M~31, presented as a function of the radius from the nucleus. Panel (a) is for the hard component, while panels (b)--(d) are for the softer thermal components with the temperature of 0.6, 0.3 and 0.1 keV, respectively.
The three lines represent exponential functions of the projected radius $r$ and the scale height $H$, in the form of $\propto e^{-r/H}$ (see text; $\S$~5).
}
\label{fig:luminosity}
\end{figure}

The surface brightness of the four components from these fits is plotted in Figure~\ref{fig:luminosity} as a function of the radius.
The results on the three diffuse X-ray components thus agree approximately within errors between the two satellites, even though the contribution from point sources is different; the ``diffuse'' spectra obtained with ${\it XMM}$-${\it Newton}$ and ${\it Chandra}$ are contributed by $\sim 15\%$ and $\sim 6\%$ from luminous ($\gtrsim 3 \times 10^{35}$ erg s$^{-1}$) sources, respectively.
The slight disagreement between them, seen in the innermost annulus, may be attributed to a much heavier point-source elimination from the ${\it XMM}$-${\it Newton}$ image.

In Figure~\ref{fig:luminosity}, all the MKL components exhibit gradually decreasing radial brightness profiles.
If we fit them with an exponential function as $\propto e^{-r/H}$, where $r$ is the projected radial distance from the nucleus and $H$ is the scale height, we obtain $H = 1\arcmin.00 \pm 0\arcmin.15$, $2\arcmin.0 \pm 0\arcmin.4$ and $1\arcmin.6 \pm 0\arcmin.3$, for the 0.6, 0.3 and 0.1 keV components, respectively.
Thus, the 0.6 keV component is considerably more centrally peaked than the other two, of which the profiles are not significantly different.

The integrated ${\it XMM}$-${\it Newton}$ luminosities of the three MKL components within $6\arcmin$ become $\sim 1.2 \times 10^{38}$ erg s$^{-1}$, $\sim 1.6 \times 10^{38}$ erg s$^{-1}$ and  $\sim 4 \times 10^{37}$ erg s$^{-1}$ in the order of decreasing temperature.
The former two values are consistent with the ${\it ASCA}$ results (Paper~1) within errors, where we identify the 0.9 keV RS component detected by ${\it ASCA}$ with the 0.6 keV MKL plasma here.

\section{DISCUSSION}

Using the ${\it XMM}$-${\it Newton}$ and ${\it Chandra}$ data, we have confirmed that the spectra of the central regions of M~31, obtained after removing point sources above $\gtrsim 3 \times 10^{35}$ erg s$^{-1}$, can be described by a sum of the four spectral components: the hard component which is mainly contributed by the photons which spill over from the excluded point sources, and the 0.6, 0.3 and 0.1 keV MKL components.
The softer three thermal components have been quantified in a consistent manner by the three satellites (${\it XMM}$-${\it Newton}$, ${\it Chandra}$ and ${\it ASCA}$).
All the thermal components are radially extended to a typical scale of $\sim$ 1$\arcmin$.5 (0.3 kpc), in agreement with the repeated detections of unresolved, possibly diffuse, soft X-ray emission
(Trinchieri and Fabbiano 1991; Primini et al.\ 1993; Supper et al.\ 1997; West et al.\ 1997; Trinchieri et al.\ 1999; Borozdin and Priedhorsky 2000; Primini et al.\ 2000; Garcia et al.\ 2000; SEA01; Paper~1; Dosaj et al.\ 2002).
Below, we discuss the nature of the four components.

\subsection{Point Sources}

With this analysis, we have reconfirmed that the integrated binary component dominates the spectra above $\sim 2$ keV, as already known previously
(Makishima et al.\ 1989b; Trinchieri et al.\ 1999; SEA01; Paper~1).
Using the ${\it XMM}$-${\it Newton}$ data, we also confirmed that, in the energy band above 0.6 keV, the summed spectra of the detected point sources cannot be reproduced by the simple PL model, while the LMXB modeling is successful when adding the 2 MKL components ($\S$~3.1).
These results agree with the main achievement of \citet{M31_Makishima} using ${\it Ginga}$ and Paper~1 using ${\it ASCA}$.

For a further confirmation of self-consistency, we attempted to reproduce the summed point-source spectra over the entire 0.4--10 keV (rather than in the $>$ 0.6 keV range) by the LMXB+3MKL model, of which the MKL temperatures were constrained at 0.6, 0.3 and 0.1 keV and the metal abundances other than nitrogen of all the three MKL components were fixed at 0.3 solar units.
The result is moderately successful with $\chi^{2}$/d.o.f.\ = 1748/1408, and the fit residuals are at most 10\% of the data.
The obtained luminosities of the 3 MKL components become $\sim 3 \times 10^{37}$ erg s$^{-1}$, $\sim 7 \times 10^{37}$ erg s$^{-1}$ and $\sim 1.6 \times 10^{37}$ erg s$^{-1}$ in the order of decreasing temperature, and correspond to $\sim$ 30--40\% of the integrated luminosity within 6$\arcmin$ of each diffuse emission ($\S$~5).
This percentage is reasonable, considering that the event-collection area of the summed point-source spectra occupies $\sim$ 10\% of the 6$\arcmin$ radius region, and that the point sources concentrate around the nucleus where the diffuse emission is also bright.
This result confirms that the 3 MKL components found in the summed point-source spectra are contaminations of the diffuse X-ray emission.

\subsection{Hotter Diffuse Emission}

The 0.5--10 keV luminosity of the hard component found in the ${\it XMM}$-${\it Newton}$ data, represented by the LMXB model, becomes $3.7 \times 10^{38}$ erg s$^{-1}$ when integrated over the $6\arcmin$ radius.
It is primarily attributable to the spill-over from the excluded point sources ($\sim 2.6 \times 10^{38}$ erg s$^{-1}$; $\S$~3.2), with a smaller contribution from unresolved faint sources ($\lesssim 5 \times 10^{37}$ erg s$^{-1}$).
After subtracting these two estimates, we are left with an unexplained hard-component luminosity of $\sim 6 \times 10^{37}$ erg s$^{-1}$ in the ${\it XMM}$-${\it Newton}$ spectra of the diffuse emission.
Similarly, the luminosity of the hard component integrated over the $3\arcmin$ radius derived from the ${\it Chandra}$ data, $1.3 \times 10^{38}$ erg s$^{-1}$, exceeds the sum of the spill-over ($\sim 4.5 \times 10^{37}$ erg s$^{-1}$) and unresolved sources ($\lesssim 1 \times 10^{37}$ erg s$^{-1}$) as described in $\S$~4.
These facts suggests the presence of yet another, rather hot, diffuse emission in the central region of M~31.

In a detailed examination of the ${\it XMM}$-${\it Newton}$ diffuse spectra (Figure~\ref{fig:0.4-7keV}), we find a hint of excess just below 7 keV, which may correspond to Fe-K lines.
Indeed, a single PL fit to the 3--7 keV portion of the ${\it XMM}$-${\it Newton}$ spectra within 6$\arcmin$, with $\chi^{2}$/d.o.f.\ = 165/159, is improved to $\chi^{2}$/d.o.f.\ = 152/157, by adding a narrow Gaussian with an equivalent width $121^{+59}_{-56}$ eV and a center energy of $6.64^{+0.07}_{-0.09}$ keV.
According to an ${\it F}$-test, the Gaussian line is significant at a $>$ 99\% confidence level.
Moreover, if we add one more MKL component to the LMXB+3MKL model in $\S$~3.3, where the metal abundances other than nitrogen of all the four MKL components are fixed at 0.3 solar units referring to Oxygen abundances of planetary nebulae in the bulge of M31 \citep{M31_abu_Jacoby}, the entire 0.4--7 keV ${\it XMM}$-${\it Newton}$ spectra in the 6$\arcmin$ radius is reproduced better ($\chi^{2}$/d.o.f.\ decreasing to 768/701 from 792/703), with insignificant changes in the parameters of the other MKL components.
The temperature and luminosity of the added fourth MKL have been obtained as $3.7^{+1.4}_{-0.7}$ keV and $(1.1\pm0.4) \times 10^{38}$ erg s$^{-1}$ respectively, within the 6$\arcmin$ radius.
The inferred luminosity of this hot MKL component agrees with the unexplained hard-component luminosity.
Although the ${\it Chandra}$ spectrum within 3$\arcmin$ does not clearly reveal the Fe-K line, the same analysis yields a hot MKL temperature of $> 5$ keV and a luminosity of $(3\pm1) \times 10^{37}$ erg s$^{-1}$.
The center energy of the Fe-K line, the implied temperature (several keV), and the luminosity of this hot and possible extended thermal emission are similar to those of the hot component of the Galactic ridge emission \citep{ridge_Koyama,ridge_Kaneda,ridge_Valinia} and the Galactic bulge emission \citep{bulge_Kokubun}.

\subsection{The Three Softer Optically-Thin Thermal Plasma Components}

The soft X-ray ($\lesssim$ 2 keV) spectra of the diffuse emission generally require more than one temperature components, even if we use relatively narrow (1$\arcmin$ $\sim$ 0.2 kpc) annular regions.
Furthermore, the three characteristic temperatures ($\sim$ 0.6, $\sim$ 0.3 and $\sim$ 0.1 keV) have been found consistently, regardless of the data accumulation region or the employed instrument.
Therefore, we regard the three temperatures as real existences, rather than caused by temperature gradients.
The larger scale radii of the softer two components than that of the 0.6 keV emission (Figure~\ref{fig:luminosity}) can explain the outward temperature decrease (from $\sim$ 0.5 keV near the nucleus to $\sim$ 0.3 keV at 8$\arcmin$), which \citet{M31_diffuse_Dosaj} obtained using a single-temperature approach.


The $\sim 0.6$ keV thin-thermal plasma emission is a new component which has first been detected in Paper~1, but not by other authors
(Primini et al.\ 2000; Garcia et al.\ 2000; SEA01; Dosaj et al.\ 2002).
Employing an improved modeling, we have successfully detected this component in the ${\it XMM}$-${\it Newton}$ and ${\it Chandra}$ data.
As argued in Paper~1, the obtained temperature and luminosity are similar to those of the soft component observed in the Galactic ridge emission and the Galactic bulge emission.


The 0.3 keV MKL component has been revealed previously by
Primini et al.\ (2000), Garcia et al.\ (2000), SEA01, Paper~1 and Dosaj et al.\ (2002).
In view of the agreement on this component among the three satellites, its reality appears to be secure.
As proposed in
West et al.\ (1997) and Paper~1,
it can be considered as an assembly of stellar coronae, although this does not exclude other possibilities.


The last diffuse X-ray plasma component, of which the temperature is $\sim 0.1$ keV, has been newly revealed in the present study.
The obtained temperature and luminosity agree between the ${\it XMM}$-${\it Newton}$ and ${\it Chandra}$ data sets.
One obvious candidate for this coolest MKL component is an assembly of hot bubbles, such as the one surrounding the solar system with a temperature $\sim 0.1$ keV \citep{hot_bubble_Tanaka,Einstein_Rosner,ROSAT_Snowden,ASCA_Gendreau}.
Using ${\it ROSAT}$, \citet{ROSAT_Snowden} estimated the local hot bubble to have a luminosity of (5--12) $\times 10^{35}$ erg s$^{-1}$ and a radius of 100--200 pc.
Then, the measured luminosity of the 0.1 keV MKL component in M~31 within 6$\arcmin$ (1.2 kpc), $\sim 3.0 \times 10^{38}$ erg s$^{-1}$ (0.1--2.4 keV), could be explained if similar bubbles are present in M~31 with a relatively high ($\sim$ 0.8) volume filling factor, assuming a disk scale height of $\sim 200$ pc.
The local hot bubble around the solar system itself contributes little to the observed 0.1 keV MKL component.

\subsection{Comparison with ${\it ROSAT}$ Image Analyses}

Now that we have confirmed the consistency among the ${\it ASCA}$, ${\it XMM}$-${\it Newton}$ and ${\it Chandra}$ data sets, we finally examine their consistency with the ${\it ROSAT}$ image analysis.
Using the ${\it ROSAT}$ images, \citet{M31_Supper} and \citet{M31_West} showed that there remains apparently ``diffuse'' emission in the central 5$\arcmin$ radius region of M~31, after subtracting detected point sources.
Its count rate was measured at 0.247 counts s$^{-1}$ without correction for the eliminated area around detected point sources \citep{M31_Supper}.
Based on our spectral analyses, the total luminosity of the truly diffuse emission is estimated as $\sim 5 \times 10^{38}$ erg s$^{-1}$ in the corresponding region and the ${\it ROSAT}$ energy band (0.1--2.4 keV).
This in turn is estimated to give $\sim$ 0.28 counts s$^{-1}$ in the ${\it ROSAT}$/PSPC (from WebPIMMS), in a rough agreement with the actually observed rate.
Thus, our results are also consistent with the ${\it ROSAT}$ data.

\clearpage

\clearpage



\begin{table*}
\caption{${\it XMM}$-${\it Newton}$ and ${\it Chandra}$ observations of the central region of M~31 utilized in the present paper.}
\label{tab:obs}
\begin{center}
\begin{tabular}{cccccc}
\hline
\hline
Satellite & Date & Obs ID & Aim Point &  \multicolumn{2}{c}{Exposure Time (ks)}\\
\cline{5-6}
          &      &        & $\alpha$/$\delta$~(J2000) & total & screened \\
\hline
${\it XMM}$-${\it Newton}$ & 2002 Jan.\ 6 & 0112570101 & ${00^{\rm h}42^{\rm m}43^{\rm s}.79}$ & 64 & 60 (MOS) \\
          &      &        & ${41\arcdeg15\arcmin36\arcsec.3}$ &  & 49 (PN) \\
\hline
${\it Chandra}$ & 2001 Oct.\ 05 & 1575 & ${00^{\rm h}42^{\rm m}44^{\rm s}.40}$ & 38 & 29 \\
          &      &        & ${41\arcdeg16\arcmin08\arcsec.30}$ &  & \\
\hline
\end{tabular}
\end{center}
\end{table*}

\begin{table*}
\caption{Model fit results to the ${\it XMM}$-${\it Newton}$ spectra summed over the point sources detected within the central $6\arcmin$ radius of M~31.$^{\rm a}$}
\label{tab:lmxb}
\begin{center}
\begin{tabular}{llccccc}
\hline
\hline
\multicolumn{2}{l}{Range} & \multicolumn{3}{c}{Hard Component} & MKL$^{\rm b}$ &  \\
\cline{3-5}
 & Model & $N_{\rm H}^{\rm c}$ & $kT_{\rm in}^{\rm d,e}$/$\Gamma^{\rm e}$ & $kT_{\rm BB}^{\rm d,e}$ & $kT_{\rm MKL}^{\rm d}$ & $\chi^2$/d.o.f.\ \\
\hline
\multicolumn{2}{l}{2--10 keV} \\
 & LMXB & $< 11$ & $0.83^{+0.06}_{-0.05}$ & $1.7\pm0.1$ & $\cdots$ & 912/871 \\
 & PL & $67^{+10}_{-9}$ & $2.03\pm0.03$ & $\cdots$ & $\cdots$ & 985/873 \\
\hline
\multicolumn{2}{l}{0.6--10 keV} \\
 & LMXB$^{\rm f,g}$ & 6.7$^{\rm h}$ & 0.7 & 2 (fix) & $\cdots$ & 4148/1340 \\
 & PL & $7.0^{+0.3}_{-0.6}$ & $1.78\pm0.01$ & $\cdots$ & $\cdots$ & 1728/1340 \\
 & LMXB+2MKL$^{\rm f}$ & 6.7$^{\rm h}$ & $0.89^{+0.02}_{-0.01}$ & 2 (fix) & $1.06\pm0.06$ & \\
 & & & & & $0.30\pm0.01$ & 1397/1336 \\
\hline
\multicolumn{6}{@{}l@{}}{\hbox to 0pt{\parbox{160mm}{\footnotesize
$^{\rm a}$ The MOS and PN spectra are fitted jointly. All the errors are single-parameter 90\% confidence limits.\\
$^{\rm b}$ The metal abundances and absorption column density of the MKL components are fixed at the 1.0 solar values and the Galactic value of $6.7 \times 10^{20}$ cm$^{-2}$, respectively. \\
$^{\rm c}$ The absorbing hydrogen column density, in the unit of $10^{20}$ cm$^{-2}$.\\
$^{\rm d}$ Temperatures are all in the unit of keV.\\
$^{\rm e}$ $kT_{\rm in}$ is the innermost temperature of the DBB model, $\Gamma$ is the PL photon index, and $kT_{\rm BB}$ is the temperature of the BB component.\\
$^{\rm f}$ The temperature of the BB component ($kT_{\rm BB}$) consisting the LMXB model is fixed at 2 keV.\\
$^{\rm g}$ Errors are not shown because of the poor fit.\\
$^{\rm h}$ The absorption column density is fixed at the Galactic value of $6.7 \times10^{20}$ cm$^{-2}$ (see text; $\S$~3.1).\\
}\hss}}
\end{tabular}
\end{center}
\end{table*}

\begin{table*}
\caption{The same as Table~\ref{tab:lmxb}, but for the diffuse X-ray emission within the central $6\arcmin$ radius of M~31.$^{\rm a,b}$}
\label{tab:fit}
\begin{center}
\begin{tabular}{llcccc}
\hline
\hline
\multicolumn{2}{l}{Range} & Hard Component & MKL & Abundances$^{\rm c}$ & \\
 & Model & $kT_{\rm in}$/$\Gamma$ &  $kT_{\rm MKL}$ & & $\chi^2$/d.o.f.\ \\
\hline
\multicolumn{2}{l}{0.6--7 keV} \\
 & LMXB+1MKL$^{\rm d}$ & $0.93\pm0.06$ & $0.39\pm0.01$ & $0.05\pm0.01$ & 953/637 \\
 & PL+1MKL & $1.80^{+0.05}_{-0.04}$ & $0.38\pm0.01$ & $0.09^{+0.02}_{-0.01}$ & 964/638 \\
 & LMXB+2MKL$^{\rm d}$ & $0.89^{+0.06}_{-0.07}$ & $0.62\pm0.03$ & $0.12\pm0.03$ & \\
 & & & $0.28^{+0.02}_{-0.01}$ & $\cdots$$^{\rm e}$ & 718/635 \\
 & LMXB+2RS$^{\rm d}$ & $1.1^{+0.2}_{-0.1}$ & $0.81^{+0.02}_{-0.03}$$^{\rm f}$ & $0.08\pm0.01$ & \\
 & & & $0.29\pm0.01$$^{\rm f}$ & $\cdots$$^{\rm e}$ & 856/635 \\
\hline
\multicolumn{2}{l}{0.4--7 keV} \\
 & LMXB+2MKL$^{\rm d}$ & $0.97^{+0.09}_{-0.08}$ & $0.58\pm0.01$ & $0.07\pm0.01$$^{\rm g}$ &  \\
 & & & $0.21\pm0.01$ & $\cdots$$^{\rm e}$ & 791/704 \\
 & LMXB+3MKL$^{\rm d}$ & $0.88^{+0.08}_{-0.07}$ & $0.61^{+0.03}_{-0.02}$ & $0.09^{+0.02}_{-0.01}$$^{\rm g}$ & \\
 & & & $0.30^{+0.02}_{-0.03}$ & $\cdots$$^{\rm e}$ & \\
 & & & $0.12^{+0.03}_{-0.02}$ & $\cdots$$^{\rm e}$ & 770/702 \\
\hline
\multicolumn{6}{@{}l@{}}{\hbox to 0pt{\parbox{160mm}{\footnotesize
$^{\rm a}$ The spectra were obtained after removing the point sources.\\
$^{\rm b}$ The absorptions of all model components are fixed at the Galactic value of $6.7 \times10^{20}$ cm$^{-2}$.\\
$^{\rm c}$ The abundances are left free, but constrained to obey the solar ratios.\\
$^{\rm d}$ The temperature of the BB component ($kT_{\rm BB}$) consisting the LMXB model is fixed at 2 keV.\\
$^{\rm e}$ Tied to the values of the hotter thermal component.\\
$^{\rm f}$ The temperature of the RS component.\\
$^{\rm g}$ The nitrogen abundance is allowed to take an independent free value (see text; $\S$~3.3).\\
}\hss}}
\end{tabular}
\end{center}
\end{table*}

\begin{table*}
\caption{Model fit results to the 0.45--7 keV ${\it Chandra}$ ACIS-S3 spectrum of the diffuse X-ray emission within the central $3\arcmin$ radius of M~31.$^{\rm a,b,c}$}
\label{tab:fit2}
\begin{center}
\begin{tabular}{llcccc}
\hline
\hline
\multicolumn{2}{l}{Model}  & Hard Component & MKL & Abundances$^{\rm d}$ \\
 & & $kT_{\rm in}^{\rm e,f}$ & $kT_{\rm MKL}^{\rm e}$ &  & $\chi^2$/d.o.f.\ \\
\hline
 & LMXB$^{\rm g}$ & 0.2 & $\cdots$ & $\cdots$ & 2321/123 \\
 & LMXB+1MKL & $1.1\pm0.2$ & $0.32\pm0.01$ & $0.08\pm0.01$ & 211/119 \\
 & LMXB+2MKL & $1.39^{+0.05}_{-0.06}$ & $0.49^{+0.03}_{-0.01}$ & $0.09\pm0.01$ & \\
 & & & $0.26\pm0.01$ & $\cdots$$^{\rm h}$ & 174/117 \\
 & LMXB+3MKL & $0.89^{+0.02}_{-0.01}$ & $0.60^{+0.03}_{-0.02}$ & $0.32^{+0.02}_{-0.01}$ & \\
 & & & $0.30\pm0.01$ & $\cdots$$^{\rm h}$ & \\
 & & & $0.10\pm0.01$ & $\cdots$$^{\rm h}$ & 149/115 \\
\hline
\multicolumn{6}{@{}l@{}}{\hbox to 0pt{\parbox{160mm}{\footnotesize
$^{\rm a}$ $N_{\rm H}$ is fixed at the Galactic value of $6.7 \times 10^{20}$ cm$^{-2}$.\\
$^{\rm b}$ The temperature of the BB component ($kT_{\rm BB}$) consisting the LMXB model is fixed at 2 keV.\\
$^{\rm c}$ All the errors are single-parameter 90\% confidence limits.\\
$^{\rm d}$ The abundances are left free, but constrained to obey the solar ratios, in contrast to the nitrogen abundance which is allowed to take an independent free value (see text; $\S$~3.3).\\
$^{\rm e}$ Temperatures are all in the unit of keV.\\
$^{\rm f}$ $kT_{\rm in}$ is the innermost temperature of the DBB model.\\
$^{\rm g}$ Errors are not shown because of the poor fit. \\
$^{\rm h}$ Tied to the values of the hotter thermal component.\\
}\hss}}
\end{tabular}
\end{center}
\end{table*}


\end{document}